\documentclass[twocolumn,showpacs,preprintnumbers,%
amsmath,amssymb,aps,floatfix]{revtex4}

\usepackage{graphicx}
\usepackage{dcolumn}
\usepackage{bm}

\newcommand{\spin} [2] {\left(\begin{matrix}#1\\#2\end{matrix}\right)}
\newcommand{\jjj}  [6] {\left(\begin{matrix}#1&#2&#3\\#4&#5&#6\end{matrix}\right)}
\newcommand{\val}      {\bm{\alpha}}
\newcommand{\vna}      {\bm{\nabla}}
\renewcommand{\vr}       {\mathrm{\mathbf{r}}}
\newcommand{\vp}       {\mathrm{\mathbf{p}}}
\newcommand{\vA}       {\mathrm{\mathbf{A}}}
\newcommand{\vF}       {\mathrm{\mathbf{F}}}
\newcommand{\hr}       {\hat{\vr}}

\begin{document}

\title{Solution of the time-dependent Dirac equation for describing 
multiphoton ionization of highly-charged hydrogenlike ions}  

\author{Yulian V. Vanne}
\author{Alejandro Saenz}%
\affiliation{%
AG Moderne Optik, Institut f\"ur Physik, Humboldt-Universit\"at
         zu Berlin, 
         Newtonstr. 15, D\,--\,12\,489 Berlin, Germany}%

\date{\today}%

\begin{abstract}
A theoretical study of the intense-field multiphoton ionization of 
hydrogenlike systems is performed by solving the time-dependent 
Dirac equation within the dipole approximation. It is shown that 
the velocity-gauge results agree to the ones in length gauge 
only if the negative-energy states are included in the time propagation. 
On the other hand, for the considered laser parameters, no significant 
difference 
is found in length gauge if the negative-energy states are included 
or not. Within the adopted dipole approximation the main relativistic 
effect is the shift of the ionization potential. A simple scaling 
procedure is proposed to account for this effect.

\end{abstract}


\maketitle

\section{\label{sec:Intro}Introduction} 

Future experiments using an electron-beam ion trap (EBIT) at the Linear
Coherent Light Source (LCLS) at Stanford and X-ray Free Electron Laser (XFEL)
at Hamburg are expected to permit the study of photoabsorption processes of 
highly
charged atomic ions in the wavelength range of 0.1 to 100\,nm with a peak
intensity up to $10^{25}$\,W/cm$^2$ or even higher. Also at the GSI (Darmstadt)
experimental investigations of highly charged ions exposed to intense laser
fields are planned within the SPARC project. 
The analysis of these experiments will require a relativistic treatment of the 
ion-laser interaction. 

Clearly, a full treatment demands to solve the time-dependent 
Dirac equation (TDDE) incorporating also the spatial dependence of the vector 
potential. Since such a treatment is very demanding, most earlier treatments 
adopted simplifications. Low-dimensional models are especially popular.  
Starting first with a one-dimensional treatment \cite{sfa:kyls97a}, 
elaborate two-dimensional calculations have been reported, e.\,g., 
in~\cite{sfa:rath97,sfa:mock04,sfa:mock08}. However, such models
can certainly 
not provide quantitative predictions and it is, at least {\it a priori}, not
even clear whether they are always qualitatively correct. Similarly to the 
non-relativistic case, also simplified ionization models like the strong-field 
approximation \cite{sfa:reis90a,sfa:fais04,sfa:klai06,sfm:fais09}, 
semiclassical tunneling theory \cite{sfa:milo02a}, 
or classical models like in \cite{sfa:keit95,sfa:gaie02,sfa:hetz09} have 
been proposed. However, in order to allow for
quantitative predictions or the validation of such simplified models a 
full-dimensional solution of the TDDE is needed. 

Very recently, a three-dimensional solution of the TDDE for hydrogenlike 
systems has been reported in \cite{sfa:sels09}. The radial solutions 
were expanded on a grid and the TDDE was either solved by a direct 
propagation on the grid or using a spectral expansion in field-free 
eigenstates. The spatial dependence of the vector potential of the carrier 
part of the laser pulse was also considered, while the one in the envelope 
was ignored. The velocity gauge was used and the importance of including 
the negative-energy (NE) states was emphasized in the case of a treatment 
beyond 
the dipole approximation, even for the considered laser parameters  
where the photon energy is insufficient to produce real positron-electron 
pairs. Based on general theoretical considerations it was conjectured that 
in length gauge the importance of the NE states may be reduced. 
On the other hand, it was concluded on the basis of the numerical results 
that within the dipole approximation the inclusion of NE states 
is not needed, even if the TDDE is solved in velocity gauge. A comparison 
of the numerically obtained ionization rates for various nuclear charges 
with the ones in non-relativistic approximation showed that, expectedly, 
increasing differences are found with increasing charge. However, it was 
concluded that the ionization rate (shown as a function of the peak value 
of the laser field) obtained within the relativistic TDDE calculation may 
be larger or smaller than the non-relativistic result. The authors found 
the higher rate easier to understand and could only speculate on possible 
reasons for the lower one.
 
A solution of the TDDE within the length gauge was reported  
more recently in \cite{sfa:pind10}. A direct time propagation on a grid was 
used. As in \cite{sfa:sels09} the 
electron-nucleus interaction is described by the unmodified non-relativistic 
Coulomb interaction. While solely the dipole approximation is adopted, 
not only results for one-electron, but also for two-electron ions are 
reported. In the latter case the electron-electron interaction is also 
described by the non-relativistic Coulomb interaction. No explicit 
discussion of the inclusion or omission of the NE states 
is given. Comparing for Ne$^{9+}$, Ne$^{8+}$, U$^{91+}$, and U$^{90+}$ 
the photoionization cross sections obtained by solving the TDDE for 10-cycle 
pulses (using a single pulse with one set of laser parameters for every 
ion) with the ones of relativistic perturbation theory, a good quantitative 
agreement is found. This result is, of course, expected, if the laser 
parameters are chosen in a way that perturbation theory is applicable, 
as was the case in \cite{sfa:pind10} where relatively low intensities 
(in relation to the ionization potentials and photon frequencies) were 
considered.   

In this work a further approach for solving the TDDE is presented. It is 
based on the spectral expansion in field-free eigenstates where the 
radial wavefunctions are expressed in a $B$-spline basis. It may be 
seen as an extension of the corresponding approaches for solving the 
non-relativistic time-dependent Schr\"odinger equation (TDSE) 
for one- and two-electron atoms in, e.\,g., 
\cite{sfa:tang90b,sfa:tang90,sfa:lamb98a,sfa:nubb08}, 
for molecular 
hydrogen \cite{sfm:awas05,sfm:pala06,sfm:vann09,sfm:sans10,sfm:vann10}, 
and for in principle arbitrary molecules within the single-active-electron 
approximation \cite{sfm:awas08,sfm:petr10a}.
While 
\cite{sfa:sels09} and \cite{sfa:pind10} considered mostly (or solely) 
one-photon ionization, an extension to multiphoton ionization is presented. 
Clearly, such calculations are very demanding as they require larger 
expansions, 
because more angular momenta are involved. Within the dipole approximation 
a systematic investigation of the importance of NE states as 
well as the convergence behavior within either length or velocity gauge is 
performed. A simple scaling relation is proposed that allows to relate 
relativistic TDDE solutions to the ones obtained with the non-relativistic 
TDSE, if both are performed within the dipole approximation. 

In this work atomic units ($e=m_e=\hbar=1$) are used
unless specified otherwise.

\section{\label{sec:Theory}Theory}

The dynamics of a highly charged atomic ion exposed to an external 
electromagnetic field is considered by means of both a relativistic and 
a non-relativistic treatment within the dipole approximation. 
The vector potential $\vA(t)$ is chosen in the 
form of an $N$-cycle $\cos^2$-shaped laser pulse that is linear-polarized 
along the $z$ axis,
\begin{equation}
\vA(t) = 
\begin{cases}
\displaystyle
\mathrm{\mathbf{e}}_{z} A_0 \cos^2\left(\frac{\pi t}{T}\right) 
\sin(\omega t),
& t \in (-\frac{T}{2},\frac{T}{2}); \\
\displaystyle
0,& \text{otherwise,}
\end{cases}
\label{}
\end{equation}
where $\omega$ is the radiation frequency and $T=2\pi N/\omega$ is the pulse
duration. Ignoring finite nuclear size effects, the interaction of an 
electron with the nucleus may be approximated by the Coulomb potential
\begin{equation}
U(r) = - \frac{Z}{r},
\label{}
\end{equation}
where $Z$ is the nuclear charge. In all calculations presented in this work 
the hydrogenlike system is initially prepared in its ground state.

\subsection{\label{subsec:RelCalc}Relativistic calculations}

The relativistic dynamics of the quantum system is governed by the 
TDDE  
\begin{equation}
i\frac{\partial \Psi(t)}{\partial t} = 
                    \left[ H^\textrm{D}_0 + V(t) \right] \Psi(t)
\label{eq:TDDE}
\end{equation}
where $H^\textrm{D}_0$ is the field-free Dirac Hamiltonian $H^\textrm{D}_0$,
\begin{equation}
H^\textrm{D}_0 = c \, \val \cdot \vp + c^2 \beta + U(r),
\label{}
\end{equation}
and the interaction with the electromagnetic
field $V(t)$ can be presented within the dipole approximation
either in the velocity (V) or length (L) gauge,
\begin{equation}
 \begin{split}
  V^\textrm{V}(t) &=   c\, \val \cdot \vA(t) = c\, \alpha_z A(t), \\
  V^\textrm{L}(t) &=     \vr  \cdot \vF(t) = z F(t).
 \end{split}
\label{}
\end{equation}
Here, $\beta$ and the components of the vector $\val$ are the Dirac matrices,
$\vp = -i\vna$ is the momentum operator, $c \approx 137$ is the speed 
of light, and $\vF(t) = - d \vA(t) /dt $ is the electric field.

As is well known (see, e.\,g., \cite{gen:gran07}), the eigenstates of 
$H^\textrm{D}_0$ can be presented as four component spinors
\begin{equation}
\Phi_{\kappa m}(\vr) = \frac{1}{r} \spin{P_{\kappa}(r)\,\chi_{ \kappa,m}(\hr)}{
                                i\, Q_{\kappa}(r)\,\chi_{-\kappa,m}(\hr)} 
\label{}
\end{equation}
where $\chi_{ \kappa,m}(\hr)$ is an $ls$ coupled spherical spinor,
$\kappa$ is the relativistic quantum number of angular momentum, related to
the orbital and total angular momenta, $l$ and $j$, as 
\begin{equation}
\kappa = \begin{cases}
- (j + 1/2) = -(l+1), & \text{for}\,\, j = l+1/2, \\
\phantom{- (}j + 1/2\phantom{)} =\phantom{-} l , & \text{for}\,\, j = l-1/2; 
\end{cases}  
\label{}
\end{equation}
and the radial functions $P_{\kappa}(r)$ and $Q_{\kappa}(r)$ are solutions
of the coupled equations 
\begin{equation}
\begin{cases}
\displaystyle 
[U(r) + c^2 - E] P_{\kappa}(r) + 
        c \left[ \frac{\kappa}{r} - \frac{d}{dr} \right] Q_{\kappa}(r) = 0 \\
\displaystyle 
[U(r) - c^2 - E] Q_{\kappa}(r) + 
        c \left[ \frac{\kappa}{r} + \frac{d}{dr} \right] P_{\kappa}(r) = 0.
\end{cases}
\label{eq:RES}
\end{equation}
For two spinor states, $\Phi_{i}$ and $\Phi_{f}$, characterized by the 
quantum numbers $\{n_i,\kappa_i,j_i,l_i,m_i\}$ and 
$\{n_f,\kappa_f,j_f,l_f,m_f\}$, respectively, the time-dependent matrix 
element $V_{fi}(t) = \langle  \Phi_{f} | V(t) | \Phi_{i} \rangle $ can be 
written as
\begin{equation}
V_{fi}(t) = \delta_{m_f,m_i}\,\delta_{|l_f-l_i|,1}\, W_{fi}\, M_{fi}(t)
\label{eq:Vfi}
\end{equation}
where 
\begin{equation}
\begin{split}
W_{fi} &= (-1)^{j_f-m_f} (-1)^{j_i+1/2} \sqrt{(2j_f+1)(2j_i+1)}\\
&\times \jjj{j_f}{1}{j_i}{-m_f}{0}{m_i}
        \jjj{j_f}{1}{j_i}{-1/2}{0}{1/2}
\end{split}
\label{}
\end{equation}
and 
\begin{equation}
M^\textrm{L}_{fi}(t) = F(t) \int dr\, r\,
\Bigl[ P_i(r) P_f(r) + Q_i(r) Q_f(r)\Bigr],
\label{}
\end{equation}
\begin{equation}
\begin{split}
M^\textrm{V}_{fi}(t) = - i c A(t) \int dr\, &
\Bigl[ (1+\Delta_{fi}) P_i(r) Q_f(r) \\
& - (1-\Delta_{fi}) Q_i(r) P_f(r)\Bigr] 
\end{split}
\label{}
\end{equation}
with $\Delta_{fi} = (-1)^{j_f-j_i}(\kappa_f-\kappa_i)$.

In order to describe both bound and continuum states, the atom is confined 
within a spherical box boundary of radius $R$. This leads to the
discretization of the continuum spectrum, whereas the 
bound states remain unmodified, if $R$ is chosen sufficiently large 
and not too highly excited Rydberg states are considered.

Introducing in a region $[0,R]$ a $B$-spline set consisting of $n+2$ 
spline functions $B_i(r)$ of order $k$, the radial functions $P(r)$ and 
$Q(r)$ can be expanded in a $B$-spline basis as
\begin{equation}
P(r) = \sum_{i=1}^{n} p_{i} B_{i+1}(r)\quad
Q(r) = \sum_{i=1}^{n} q_{i} B_{i+1}(r)
\label{}
\end{equation}
where the first and the last spline are removed from the expansion
to ensure the boundary conditions $P(0) = Q(0) = P(R) = Q(R) = 0$.
Defining the coefficient vector 
\begin{equation}
 \mathbf{C} = (p_1,q_1,p_2,q_2,\dots,p_n,q_n),
\label{}
\end{equation}
the system of equations~(\ref{eq:RES}) is transformed into 
a $2n \times 2n$ generalized banded eigenvalue 
problem which is efficiently 
solved using LAPACK routine DSBGVX. This procedure yields for every value of 
$\kappa$ exactly $n$ negative energy solutions and $n$ positive energy solutions. 

As has been discussed in literature since the first solution of the
time-independent (stationary) Dirac equation using $B$ 
splines~\cite{bsp:john88}, there is the problem of the occurrence of spurious 
states \cite{gen:drak81} that are non-physical. Different procedures 
were proposed 
to avoid this problem \cite{bsp:shab04,bsp:belo08,bsp:gran09}. While there 
can be a problem with their identification in the case of many-electron
systems, it is the lowest positive energy state for $\kappa >0$ that represents a spurious 
state in the present case.

\subsection{\label{subsec:NonrelCalc}Non-relativistic calculations}

The non-relativistic TDSE

\begin{equation}
i\frac{\partial \Psi(t)}{\partial t} = 
                    \left[ H_0 + V(t) \right] \Psi(t)
\label{eq:TDSE}
\end{equation}
where $H_0$ is the non-relativistic field-free Hamiltonian $H_0$,
\begin{equation}
H_0 = \frac{\vp^2}{2} + U(r),
\label{}
\end{equation}
and in contrast to the relativistic case, the interaction with 
the electromagnetic field $V(t)$ is given in  
the velocity gauge by
\begin{equation}
  V^\textrm{V}(t) =  \vp \cdot \vA(t) = p_z A(t).
\label{}
\end{equation}
Similar to the relativistic case, the eigenstates of $H_0$, 
$\Phi_{l m}(\vr) = r^{-1} R_{l}(r) Y_{lm}(\hr)$,
are obtained by projecting the radial function $R_{l}(r)$ onto the 
$B$-spline basis
\begin{equation}
 R(r) = \sum_{i=1}^{n} \rho_i\, B_{i+1}(r)
\label{}
\end{equation}
and transforming the radial Schr\"odinger equation into the $n \times n$
generalized banded eigenvalue problem with respect to a coefficient vector
\begin{equation}
 \mathbf{C} = (\rho_1,\rho_2,\dots,\rho_n),
\label{}
\end{equation}
which yields $n$ solutions for every orbital quantum number $l$.

For two states, $\Phi_{i}$ and $\Phi_{f}$, characterized by the 
quantum numbers $\{n_i,l_i,m_i\}$ and 
$\{n_f,l_f,m_f\}$, respectively, the time-dependent matrix 
element $V_{fi}(t) = \langle  \Phi_{f} | V(t) | \Phi_{i} \rangle $ can
also be written in form of Eq.~\ref{eq:Vfi}, but with
\begin{equation}
\begin{split}
W_{fi} &= (-1)^{l_f-m_f} (-1)^{l_f} \sqrt{(2l_f+1)(2l_i+1)}\\
&\times \jjj{l_f}{1}{l_i}{-m_f}{0}{m_i}
        \jjj{l_f}{1}{l_i}{0}{0}{0}
\end{split}
\label{}
\end{equation}
and  
\begin{equation}
M^\textrm{L}_{fi}(t) = F(t) \int dr\, r\, R_i(r) R_f(r), 
\label{}
\end{equation}
\begin{equation}
M^\textrm{V}_{fi}(t) = - i A(t) \int dr\,
R_f(r) \Bigl[ \frac{l}{r} R_i(r) + R'_i(r) \Bigr].
\label{}
\end{equation}

\subsection{\label{subsec:TimeProp}Time propagation}

Within the spectral approach, the integration of the TDDE~(\ref{eq:TDDE}) 
and the TDSE~(\ref{eq:TDSE}) is performed by expanding the function 
$\Psi(t)$ describing the 
dynamics of the system in the basis of field-free eigenstates $\Phi_{K}$,
\begin{equation}
\Psi(\vr,t) = \sum_{K} C_{K}(t) e^{-i E_{K} t} \Phi_{K}(\vr) 
\label{eq:PsiExp}
\end{equation}
where the compound index $K$ represents the full set of quantum numbers 
and the coefficients $C_{K}(t=-T/2)$ are set to zero for all states except for 
the initial state for which its value is set to 1. Since the quantum number 
$m$ is conserved for both cases and the ground state is chosen as initial 
state, the value of $m$ is fixed to $1/2$ for the relativistic case and to 
$0$ for the non-relativistic case. 

Substituting Eq.~(\ref{eq:PsiExp}) into Eq.~(\ref{eq:TDDE}) or
TDSE~(\ref{eq:TDSE}) the latter ones 
are reduced to a system of coupled first-order ordinary
differential equations,
\begin{equation}
 C'_{K}(t) = \sum_{K'} e^{i (E_{K}-E_{K'}) t}\,  V_{KK'}(t) C_{K'}(t)\quad ,
\label{}
\end{equation}
which is integrated numerically using a variable-order, variable-step Adams
solver. 

The ionization yield (after the pulse) is then defined as
\begin{equation}
Y_{\rm ion} = \sum_{K} \left| C_{K}(t=T/2) \right|^{2}
\label{eq:Yion}
\end{equation}
where the summation in Eq.~(\ref{eq:Yion}) is performed over all 
discretized continuum states. The convergence can be controlled by varying 
the value of $l_\textrm{max}$ (or $j_\textrm{max}$) which
limits the number of different symmetries involved in the 
summation~(\ref{eq:PsiExp}). 

In the case of the TDDE the question of a proper treatment of the 
spurious states mentioned in Sec.\ref{subsec:RelCalc} arises. 
The success of the spectral approach relies on the completeness of 
the states included in the summation, at least for a given box size. 
In fact, already in the case of the non-relativistic Schr\"odinger 
equation the use of a finite box leads to the occurrence of 
non-physical pseudo states (see, e.\,g., \cite{sfa:lamb98a}). In this 
case it turned out to be in fact important to include those states 
in the spectral expansion, since otherwise some relevant part of the 
Hilbert space is omitted. In order to decide on the proper treatment 
of the spurious states of the Dirac equation (due to the $B$-spline 
expansion), a careful check of the relativistic sum rules~\cite{gen:drak81} 
was performed that also served as a check of the proper implementation 
of the length and velocity forms of the dipole-operator matrix elements. 
On this basis it was concluded that the spurious states could be omitted 
from the basis for solving the TDDE, since the sum rules were fulfilled 
in this case. The agreement of the TDDE solutions obtained within the 
length and the velocity gauge discussed below is another indication 
that the omission of the spurious states is appropriate.

\subsection{\label{subsec:Zvar}Computational details and scaling of the 
TDSE with $Z$}

For the sake of consistency, the same $B$-spline basis set is 
used for both the relativistic and the non-relativistic treatment. Typically, 
the values $k=9$ and $n=500$ are used to construct an almost
linear knot sequence in which the first 40 intervals increase with a geometric
progression using $g=1.05$ and all following intervals have the length of the 
40th one. Such a choice ensures an accurate numerical description in the 
vicinity of the nucleus and, at the same time, a sufficient completeness for a
description of the discretized continuum \cite{bsp:vann04}. 
Depending on the nuclear 
charge $Z$, a box size of $R=(250/Z)\,a_0$ is adopted. This choice of $R$
reflects the well-known scaling property of the time-independent Schr\"odinger 
equation of hydrogenlike systems. With the substitution $\vr'=\vr/Z$ and 
$E'=Z^2 E$ for the position $\vr$ and the energy $E$, respectively, 
the eigensolutions for a hydrogenlike system with charge $Z$ 
reduces to the one of the hydrogen atom with $Z=1$. 

Within the dipole approximation this property is preserved also for the 
TDSE, if the pulse parameters are also properly scaled, e.\,g., the time as
$t' = t/Z^2$, the laser frequency as $\omega' = Z^2 \omega$, the 
amplitude of the vector potential as $A_0' = Z A_0$, and the laser peak 
intensity as $I'=Z^6 I$. Since this property does not persist in case of 
the relativistic treatment, and the properly scaled solutions of the TDDE 
are thus not identical anymore, any deviation from the non-relativistic 
prediction based on the scaling relations can be classified as a relativistic 
effect, cf.\ \cite{sfa:sels09}.

\section{\label{sec:Res}Results and discussion}

\subsection{\label{subsec:Conv}Convergence behavior}

%
\begin{figure}[!tbp]
\begin{center}
\includegraphics[width=0.45\textwidth]{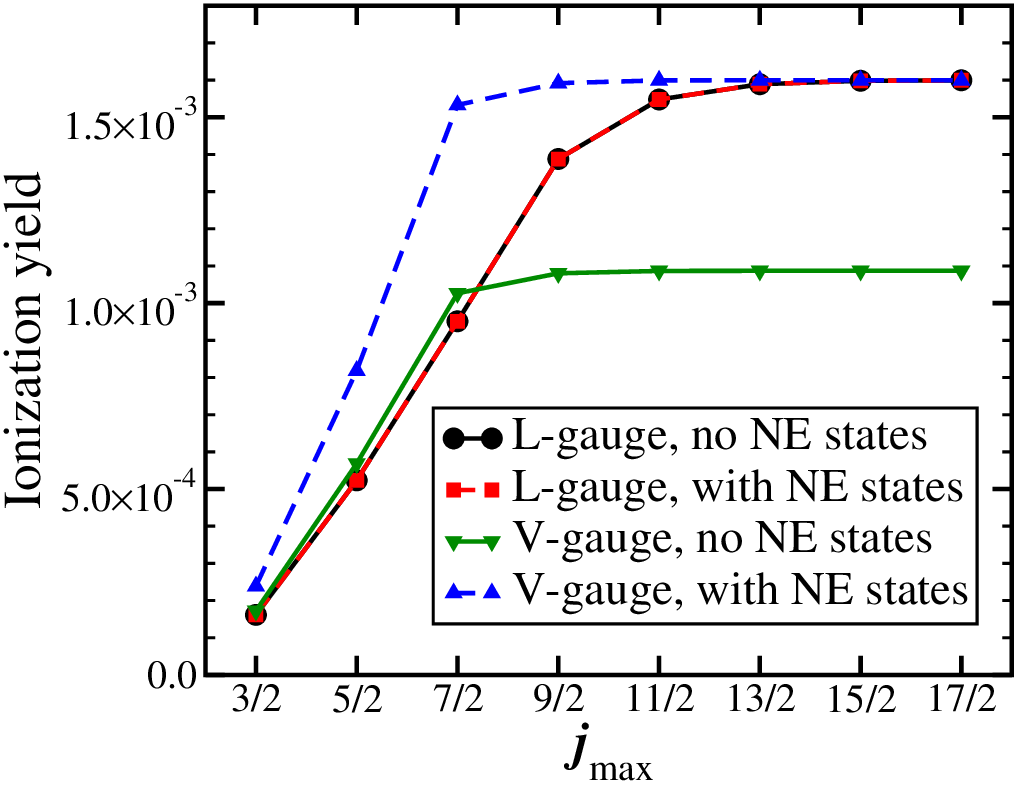}%
\caption{\label{fig:Jconv} 
Convergence study with respect to the maximal value of the total angular 
momentum $j_\textrm{max}$. The relativistic calculations 
of the ionization yield for an ion with the nuclear charge $Z=50$ exposed
to a $20$-cycle cos$^2$-shaped laser pulse with peak intensity 
$I=5\times 10^{23}$W/cm$^2$ and a photon energy of $500$ a.u.\ are 
performed using the length (L) or velocity (V) gauge 
either including or excluding the negative energy (NE) states in 
the expansion~(\ref{eq:PsiExp}).} 
\end{center}
\end{figure}
%

The handling of the NE solutions of the Dirac equation is an important issue
and steered considerable attention in the 
literature, see \cite{gen:furs08, sfa:sels09} and references therein. 
The question is whether the NE
states should be removed from the basis as long as the creation of real
positron-electron pair is energetically out of reach. Such a situation
occurs, for example, in case of ionization of an ion with the nuclear charge
$Z=50$ exposed to a pulse with photon energy 500 a.u.\ (see 
Fig.~\ref{fig:Jconv}). Whereas only 3 photons are sufficient for ionization,
more than 70 photons are required for positron-electron pair creation. 
As is demonstrated in Fig.~\ref{fig:Jconv}, the answer to the question 
of the proper treatment of the NE states depends already within  
the dipole approximation adopted here clearly on the gauge and thus 
the interaction operator used in the time propagation. For the length 
gauge the exclusion of NE states has virtually no effect on the final result. 
In contrast, for the velocity gauge the results are 
obviously different. If the NE states are included in the time propagation, 
the converged result agrees with the one obtained using the length gauge, 
whereas without the NE states the converged result differs in this example 
by a factor of about 1.5. Therefore, these states are absolutely necessary for 
obtaining the correct result. 

From the practical point of view the finding is also interesting, since 
the omission of the NE states and thus half of the total number of states 
reduces the numerical efforts tremendously. Clearly, doubling the number 
of states leads to an increase of the number of operations per time 
step by a factor 4, since the number of transition dipole matrix elements 
increases by this factor.  In the present example 
(and for a given number of $j_\textrm{max}$) the time propagation 
in length gauge without the NE states is by about a factor of 6 faster 
compared to the one where the NE states are included. The additional 
time difference (factor of about 1.5) arises from an increased number 
of time steps required for convergence in the used adaptive time 
propagation, if the NE states are included. 
In fact, the length-gauge time propagation by itself is found to be 
about 6 times faster than the one performed in velocity gauge, even if the 
NE states are included in both of them. This factor is due to the finer 
time grid needed for convergence in velocity gauge. Considering both 
effects together, even a speed-up by a factor of 50 is found when comparing 
the length-gauge calculation without NE states and the velocity-gauge 
variant with NE states that for sufficiently large value of $j_\textrm{max}$ 
both yield practically identical results.
 
However, Fig.~\ref{fig:Jconv} reveals also that the calculations in velocity 
gauge converge faster with respect to the quantum numbers $j$ included 
in the calculation. Therefore, the efficiency gain of the length gauge 
is smaller than the value given above. In the concrete example shown 
in Fig.~\ref{fig:Jconv} the (within better than 0.1\,\%) converged 
length-gauge calculation ($j_\textrm{max}=15/2$, without NE states) 
is by a factor of about 40 faster than the (within about 10$^{-3}\,$\%) 
converged velocity-gauge result ($j_\textrm{max}=11/2$, with NE states). 
Even compared to the only about $0.5\,$\% converged velocity-gauge result 
with $j_\textrm{max}=9/2$ (with NE states) there is still a factor of 
about 30 in time gain.  On the other hand, the question of the 
most efficient choice of the gauge 
depends on the laser parameters. If few-photon processes, in the most 
extreme case one-photon ionization at low intensities, are considered, 
the average angular momentum $j$ transferred to the ion is small. In 
such a case the need for a larger value of $j_\textrm{max}$ in the 
length-gauge calculation could even overcompensate the gain from the 
exclusion of the NE states. However, if the average number of absorbed 
photons increases (due to a smaller ratio of the photon frequency 
with respect to the ionization potential or due to a higher laser 
intensity), the relative increase in $j$ due to the slower convergence 
leads to a smaller increase of the total number of states than the 
inclusion of the NE states. In this multiphoton regime calculations 
in length gauge appear to be much more efficient. This finding for the 
TDDE solution seems to differs from the one for the non-relativistic 
TDSE where the velocity gauge is often supposed to converge faster 
than the length gauge. 

For extremely high intensities around and above the critical field 
strength $F_{\rm cr}=c^3\approx 2.57\,\times\,10^{6}$ where real pair 
production is possible, the inclusion of the NE states is, of course, 
required also in length gauge. In fact, already for the parameters 
discussed in the context of Fig.~\ref{fig:Jconv} we find a relative 
deviation between otherwise converged length-gauge results with 
and without NE states of about $2.2\,\times\,10^{-5}$. This 
appears reasonable, since it is of the order of the (reciprocal) 
rest energy $c^{-2} \approx 5.3\,\times\,10^{-5}$. 
For very high laser intensities the gain 
from omitting the NE states is thus lost and future calculations 
will have to show whether in that regime length-gauge calculations 
can still profit from the need of a sparser time grid, or whether 
the faster convergence with $j$ persists in velocity gauge and may 
make this gauge more efficient.

%
\begin{figure}[!tbp]
\begin{center}
\includegraphics[width=0.45\textwidth]{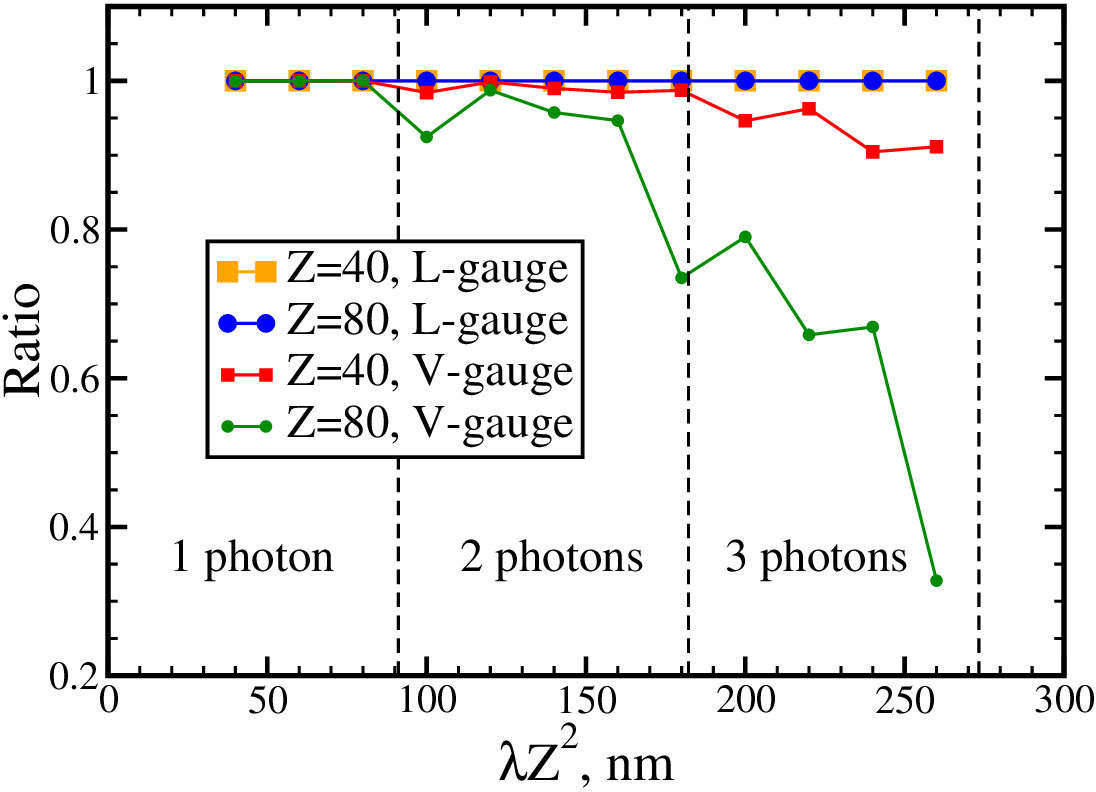}%
\caption{\label{fig:RPES} 
Ratio of the ionization yields that are obtained by either excluding or 
including the negative energy states in the 
expansion~(\ref{eq:PsiExp}) adopting either length (L) or velocity (V) gauge. 
The calculations are performed for an ion with two different values of 
nuclear charge $Z$ by varying the wavelength $\lambda$ of a $20$-cycle 
cos$^2$-shaped laser pulse with the peak intensity 
$Z^6\times 10^{13}$W/cm$^2$. The character of the ionization process
changes from single-photon ionization ($\lambda Z^2 = 40\,$nm) to
three-photon ionization ($\lambda Z^2 = 260\,$nm).
} 
\end{center}
\end{figure}
%

In \cite{sfa:sels09} it was concluded on the basis of the numerical results 
that the inclusion of the NE states is not crucial for solving the TDDE, 
if the dipole approximation is adopted. This result was found despite
the fact that the calculation was performed in velocity gauge. Therefore, 
this finding appears to contradict our conclusions. However, it turns out 
that the importance of the inclusion of the NE states depends in fact 
on the character of the multiphoton ionization process, i.\,e.\ on the 
number of photons. Figure~\ref{fig:RPES} shows the ratio of the ionization 
yields obtained either with or without inclusion of the NE states for 
both gauges and two values of the nuclear charge $Z$. In agreement with 
the findings discussed above, the inclusion of the NE states has 
practically no influence for the length-gauge results, independently of 
the number of photons involved or the nuclear charge. This changes, 
however, if the velocity gauge is used in the time propagation. 
The importance of the NE states increases with the order of the 
multiphoton process (the number of photons required for ionization) 
and with the nuclear charge. Within the one-photon regime there 
is practically no influence of the NE states.
Thus the finding in \cite{sfa:sels09}  
is confirmed that even in velocity gauge an inclusion of the NE 
states is not required for calculations in the dipole approximation. 
However, it is to be understood that those findings apply only 
to the case where one photon is sufficient to reach into the 
ionization continuum.  
Quite remarkable is also the strong dependence of the importance of the 
NE states on $Z$.  For example, for the wavelength 
$\lambda = 260/Z^2$\,nm the exclusion of the NE states leads to a three 
times smaller value of the ionization yield for $Z=80$, whereas for 
$Z=40$ the decrease is only about 10\,\%. The $Z$ dependence can  
also explain the negligible effect of NE states within the dipole 
approximation found in Fig.\,1 of \cite{sfa:sels09} where $Z=1$ was 
used. Furthermore, only the probability of remaining in the initial 
ground state was considered and this quantity is found to converge 
much faster than, e.\,g., the ionization yield. This could possibly 
make it also less sensitive to the NE states.

\subsection{\label{subsec:1ph}Single-photon ionization}

\begin{figure}[!tbp]
\begin{center}
\includegraphics[width=0.45\textwidth]{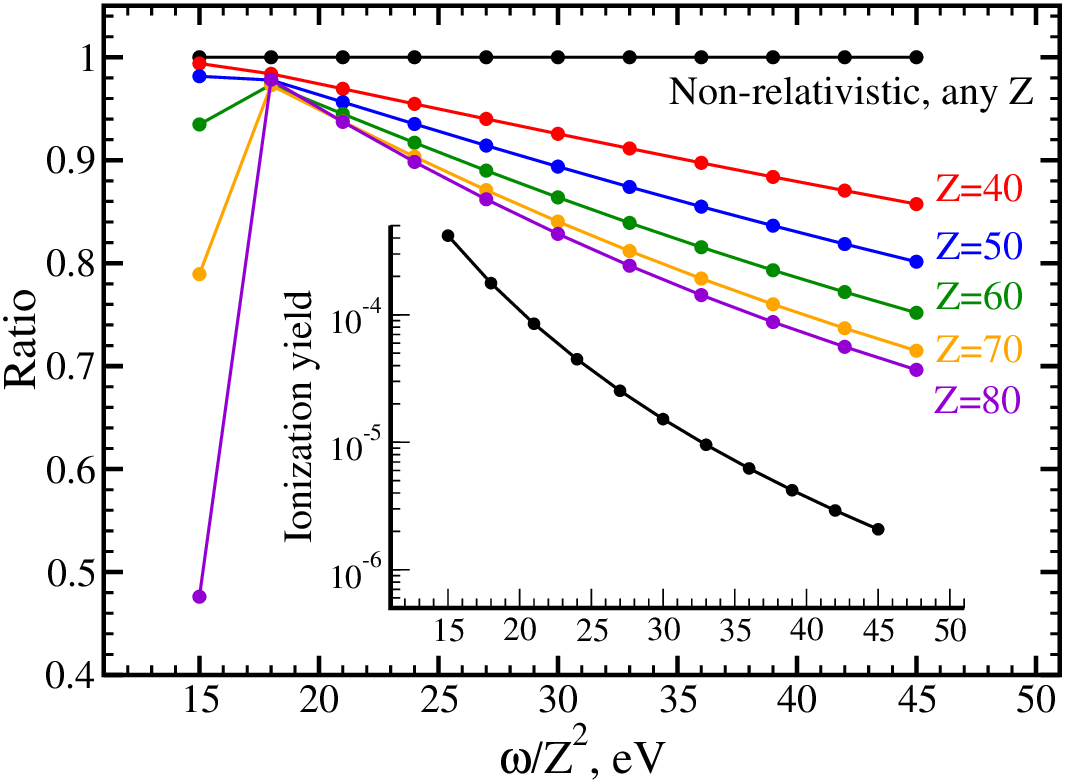}%
\caption{\label{fig:R1ph} 
Ratios of the relativistic ionization yields to the non-relativistic
ones (shown in the insert) obtained for 5 different values of 
the nuclear charge $Z$ by varying the carrier frequency $\omega$ of a 
$20$-cycle cos$^2$-shaped laser pulse with the peak intensity 
of $Z^6\times 10^{11}$W/cm$^2$.} 
\end{center}
\end{figure}

As has been discussed in Sec.~\ref{subsec:Zvar}, the solution of the TDSE 
for different nuclear charges $Z$ gives identical ionization yields, if the 
laser pulse parameters are scaled properly. For example, the insert in 
Fig.~\ref{fig:R1ph} shows the non-relativistic ionization yields obtained 
for an ion with the nuclear charge $Z$ exposed to a $20$-cycle cos$^2$-shaped 
laser pulse with a peak intensity of $Z^6\times 10^{11}$W/cm$^2$ for photon 
energies varying in the range between $15\,Z^2\,$eV and $45\,Z^2$\,eV. 
Since the ionization potential of the ion is equal to $13.6\,Z^2$\,eV, 
the ionization should occur via absorption of a single photon. 
In order to study the relativistic effects in this one-photon 
ionization regime, the TDDE is solved 
for the same system and for 5 different values 
of $Z$ (in between 40 and 80 with a stepsize of 10). 

%
%

The ratio of the 
relativistic to the non-relativistic ionization yield is shown in 
Fig.~\ref{fig:R1ph}. Two features can be observed. First, above 
$18\,Z^2\,$eV the ratio decreases with increasing photon energy. 
Thus relativistic effects are more pronounced for higher 
photon frequencies $\omega$. As could be expected, this effect becomes   
more and more important as $Z$ increases. Second, a discontinuity 
develops with increasing $Z$ at a photon energy of about $18\,Z^2\,$eV. 
Especially for large values of $Z$ the ratio 
for the lowest considered photon energy drops down substantially for  
increasing $Z$. This is a manifestation of the fact that the 
ionization potential of the ion increases as a consequence of the 
relativistic velocity of an electron in the vicinity of strong Coulomb
potentials. The shift of the ionization potential 
\begin{equation}
 \Delta I_p = I^\textrm{rel}_p(Z) - I_p^\textrm{nr}(Z) = 
      c^2\left(1- \sqrt{1-\frac{Z^2}{c^2}}\right) - \frac{Z^2}{2}
\label{}
\end{equation}
rapidly increases with increasing $Z$. (For example,  
$\Delta I_p \approx Z^4/(8c^2)$ for small $Z$.) 
Thus, for $Z=80$ the absorption of a single photon with the 
energy $15\,Z^2$\,eV is not sufficient for ionization anymore and the 
character of the ionization process changes from single-photon to two-photon
ionization. For the considered intensities the latter process 
possesses of course 
a much smaller probability. In fact, the ratio would be even smaller, 
if the ratio at $15\,Z^2$\,eV would correspond to pure two-photon ionization. 
The finite spectral width of the adopted 20-cycle laser pulse allows, 
however, one-photon ionization to occur even at this energy and increases 
thus the ionization yield. Therefore, a finer photon-frequency grid in between 
$15\,Z^2$\,eV and $20\,Z^2$\,eV would show a much smoother behavior than 
the one visible from Fig.~\ref{fig:R1ph}.  

In \cite{sfa:sels09} (Fig.\,5) relativistic effects were considered by a 
comparison of the ionization rates obtained with the TDSE and the TDDE 
where the latter was solved for 8 different values of $Z$. However, there 
the behavior was studied as a function of the laser field amplitude 
(also scaled by the corresponding non-relativistic scaling relations) 
and for a fixed photon frequency. The authors discuss stabilization, since 
the ionization rate does not increase monotonously with the field strength, 
but instead decreases for intensities beyond about $F_0=1\,Z^3$. The 
for large values of $Z$ increasing value of $F_0$ for which the ionization 
rate has its maximum is then explained by the stabilization criterion 
of Gavrila (see \cite{sfa:gavr02} and references therein) 
and the conjecture that due to the lower 
energy of the ground state in the relativistic case the condition for 
the occurrence of stabilization shifts to higher field strengths.

\begin{figure}[!tbp]
\begin{center}
\includegraphics[width=0.45\textwidth]{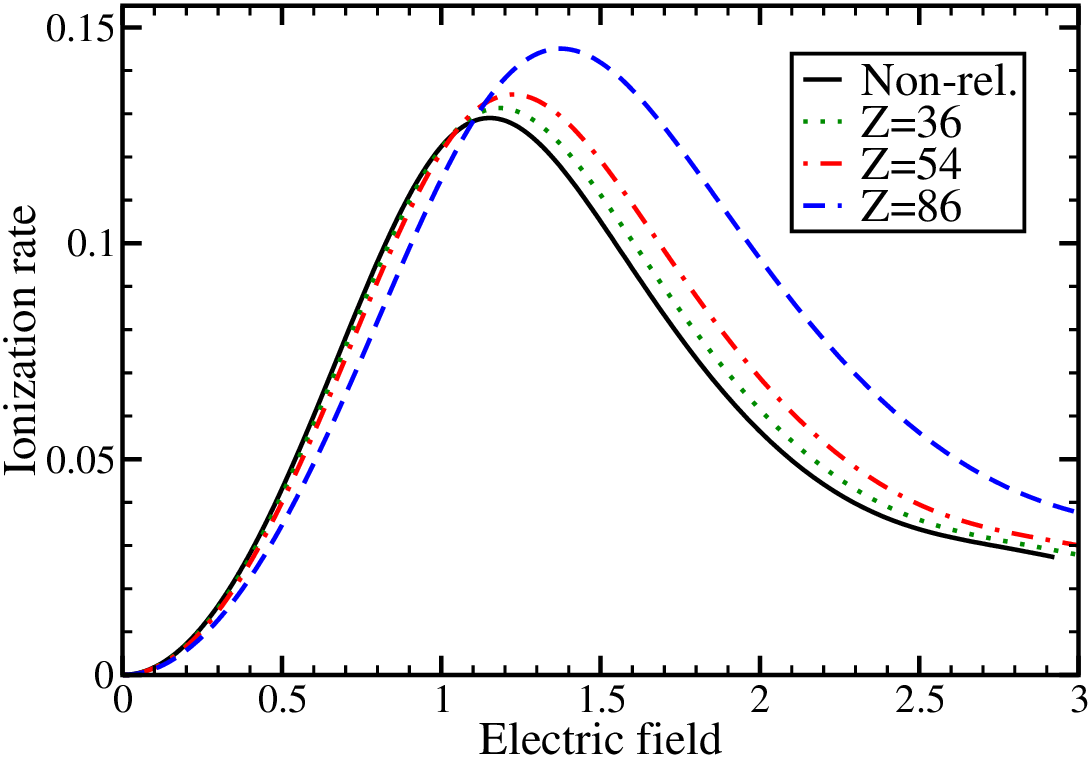}%
\caption{\label{fig:Floq} 
Ionization rate (in units of $Z^2$ a.u.) of a hydrogenlike ion  
exposed to a monochromatic laser field with the frequency $\omega = Z^2$ 
as a function of the amplitude of the electric field 
(given in units of $Z^3$ a.u.) The black solid curve presents the 
(non-relativistic) Floquet ionization rate, whereas the other curves present 
the ionization rates obtained using the scaling relation~\ref{eq:Ratesc} 
in order
to estimate the relativistic ionization rates for three different values of
$Z$.
} 
\end{center}
\end{figure}

In fact, the dependence of the ionization 
rates on $Z$ discussed in \cite{sfa:sels09} may be quantitatively understood 
from the scaling relations together with the lowering of the ground-state 
energy due to relativistic effects. This is illustrated in the 
following way. From the condition $I^\textrm{rel}_p(Z) = 
I_p^\textrm{nr}(Z')$ one finds a scaled nuclear charge%
\begin{equation}
Z'  = \sqrt{2c^2 \left(1- \sqrt{1-Z^2/c^2}\right)},
\label{eq:Zp}
\end{equation}
for a given true charge $Z$. This allows to estimate the 
relativistic ionization rate from a non-relativistic calculation 
using
\begin{equation}
\Gamma'(F_0) = \left(\frac{Z'}{Z}\right)^2\, 
                \Gamma\left( F_0 \left[\frac{Z}{Z'}\right]^3 \right) \quad ,
\label{eq:Ratesc}
\end{equation}
if the dependence of the non-relativistic ionization rate on the 
photon frequency is ignored. In order to obtain non-relativistic 
ionization rates $\Gamma(F_0)$, Floquet calculations 
were performed as a function of the peak amplitude 
$F_0$ of the electric field using program STRFLO~\cite{sfa:potv98}. 
They are shown in Fig.~\ref{fig:Floq} and are in reasonable agreement 
with the TDSE rates in~\cite{sfa:sels09}. Based on the non-relativistic 
Floquet rates the scaling relation (\ref{eq:Ratesc}) allows to 
estimate the relativistic rates that are also shown in Fig.~\ref{fig:Floq} 
for $Z=36$, 54, and 86. The qualitative agreement with the TDDE results 
in Fig.\,5 of \cite{sfa:sels09} is satisfactory and especially 
the shift of the maximum to higher fields is well reproduced. 
However, in agreement with the discussion in \cite{sfa:sels09} 
the model predicts that the maximum of the ionization rate 
increases monotonically with $Z$. This is in contrast to the 
numerical findings in \cite{sfa:sels09}.      

%
%

%
\begin{figure}[!tbp]
\begin{center}
\includegraphics[width=0.45\textwidth]{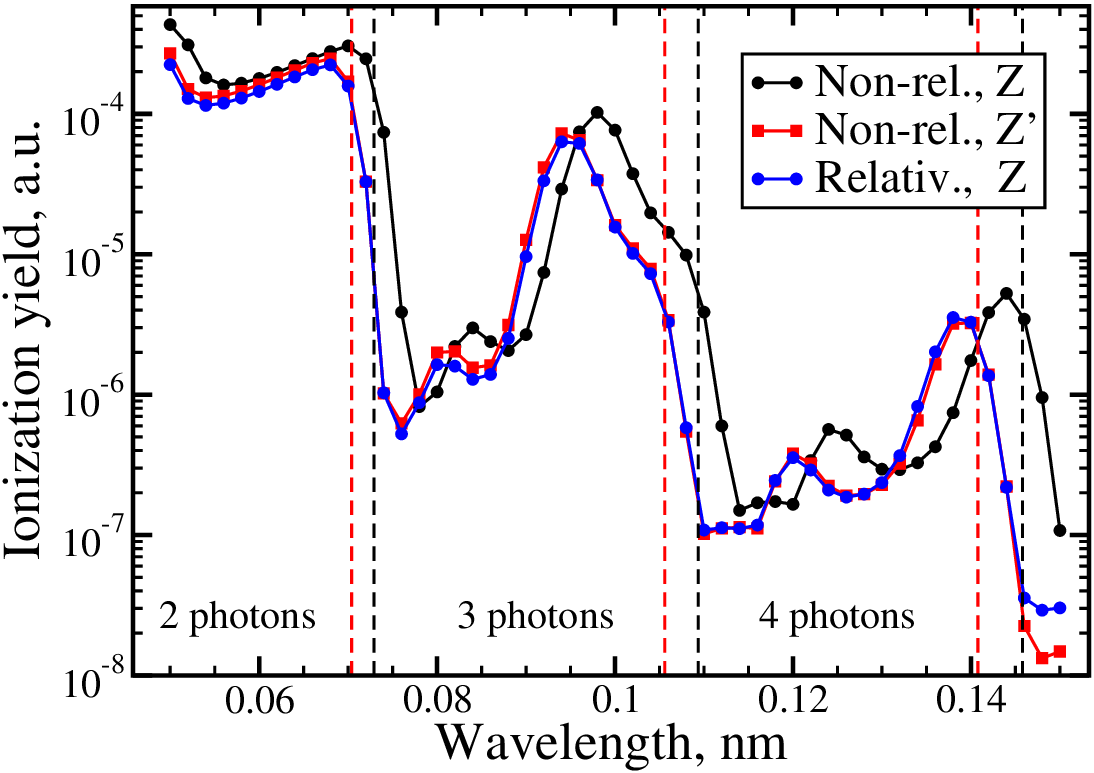}%
\caption{\label{fig:MPI} 
Relativistic (blue circles) and non-relativistic (black circles)
ionization yields obtained for an ion with the nuclear charge $Z=50$ 
exposed to a $20$-cycle cos$^2$-shaped laser pulse with a peak intensity 
of $5\times10^{22}$ W/cm$^2$ and various laser wavelengths. Additionally,
the non-relativistic ionization yields (red squares) for an ion 
with the nuclear charge $Z'=50.88$ are shown whose non-relativistic 
ionization potential is equal to the relativistic ionization potential of 
the ion with $Z=50$. The non-relativistic $N$-photon thresholds ($N = 2-4$) 
for $Z$ and $Z'$ are indicated by black and red vertical dashed lines,
respectively.} 
\end{center}
\end{figure}
%

\subsection{\label{subsec:MPI}Multiphoton ionization}

If many photons are required for ionization, the relativistic results
may deviate from the non-relativistic ones by an order of magnitude or more. 
This is demonstrated in Fig.~\ref{fig:MPI} where relativistic and 
non-relativistic ionization yields are compared for an ion with the nuclear
charge $Z=50$. The laser wavelength varies in the range from 0.05\,nm
(2-photon ionization) to 0.15\,nm (5-photon ionization). Whereas the general
structure of the wavelength dependence is similar for both relativistic and 
non-relativistic treatments, the difference in the positions of all pronounced 
features like peaks and minima may result in a substantial
discrepancy, if the ion yields are compared at a single photon frequency. 
Obviously, the sharp steps are due to the $N$-photon channel closings 
and the peaks are signatures of resonantly enhanced multiphoton ionization 
(REMPI) processes. Their positions are directly related to excitation 
energies and the ionization potential, respectively, and both 
are higher in the relativistic than in the non-relativistic case. 

In view of the scaling relation (\ref{eq:Zp}) it is interesting 
to compare the TDDE result (for $Z=50$) with the non-relativistic 
TDSE result for the scaled nuclear charge $Z'= 50.88$. As can be seen 
from  Fig.~\ref{fig:MPI}, the obtained TDSE results deviate from the
relativistic ones for $Z=50$ only by a few percent. This indicates 
that the change of the ionization potential is the by far dominating 
relativistic effect, at least if multiphoton ionization is described within 
the dipole approximation. Other possible effects like the splitting 
of (resonant) intermediate states due to spin-orbit coupling are 
thus very small, but can be quantified more easily, after the 
scaling relation has accounted for the main effect. 
The dominant influence of the shift of the ionization potential is 
likely to depend on the considered laser intensity 
and photon frequency. It should be reminded that the Keldysh parameter 
$\gamma=\sqrt{2\,I_p}\omega/F_0$ \cite{sfa:keld65} 
varies in Fig.~\ref{fig:MPI} in between 38.17 and 12.72. 
This is deep in the multiphoton regime, in fact even in the perturbative 
one.


\section{\label{sec:Concl}Conclusion}

Single- and multiphoton ionization of highly charged atomic ions
has been numerically studied by a direct solution of the time-dependent Dirac
equation within the dipole approximation. The stationary Dirac equation is 
solved by projecting the radial part onto a $B$-spline basis and the 
obtained field-free eigensolutions are used in the subsequent time 
propagation. Results for both length and velocity
gauges for describing the ion-field interaction are obtained and compared. 
The inclusion of the negative-energy Dirac states for the description of 
the relativistic dynamics is shown to be important in the case of the 
multiphoton ionization, if the velocity gauge is adopted, even if the 
considered intensities and frequencies are too low for allowing 
non-negligible real pair 
creation. If ionization occurs via absorption of a single
photon or the time propagation is performed in the length form, the role
of negative energy states is much less significant. 

Comparing solutions of the time-dependent Dirac and Schr\"odinger equations
for the same ion and laser pulses, the relativistic change of the  
ionization potential is demonstrated to dominate other relativistic effects
in the case of multiphoton ionization. It is shown that this effect 
is successfully accounted for by a simple scaling relation.

\section*{Acknowledgments}
The authors acknowledge financial support by the German Ministry for Research 
and Education (BMBF) within the SPARC.de network under grant 06BY9015,   
by the COST programme CM\,0702, and by the Fonds der Chemischen Industrie. 
This work was supported in parts by the National Science Foundation under 
Grant No.\ NSF PHY05-51164.


\bibliographystyle{apsrev}


\end{document}